\documentclass[aps,pra,twocolumn,showpacs,letterpaper,tighten,float,superscriptaddress,nofootinbib]{revtex4-1}
\usepackage{amssymb}
\usepackage{amsbsy}
\usepackage{amsmath}
\usepackage{mathrsfs}
\usepackage{epsfig}
\usepackage{graphicx}
\usepackage{array}
\usepackage{textcomp}
\usepackage{color}
\usepackage{braket}
\usepackage{bm,hyperref}
\usepackage[justification=raggedright,font=small]{caption}
\usepackage[titletoc,toc,title]{appendix}
\usepackage[normalem]{ulem}
\usepackage{cancel}

\usepackage[labelformat=simple]{subcaption}

\definecolor{myblue}{rgb}{.93, .93, 1}

\definecolor{darkgreen}{rgb}{0,0.7,0}

\newcommand{\beq}{\begin{equation}}
\newcommand{\eeq}{\end{equation}}

\newcommand{\bpm}{\begin{pmatrix}}
\newcommand{\epm}{\end{pmatrix}}
\newcommand{\bmm}{\begin{matrix}}
\newcommand{\emm}{\end{matrix}}

\graphicspath{{./Figures/}}

\begin{document}
\title{On The Low Speed Limits of Lorentz's Transformation\\ - How relativistic effects retain or vanish in electromagnetism}

\author{Hao Chen}
\email{chen.hao@princeton.edu}
\affiliation{Department of Electrical and Computer Engineering, Princeton University, Princeton, New Jersey 08544, USA}
\affiliation{Department of Physics, Princeton University, Princeton, New Jersey 08544, USA}

\author{Wei E. I. Sha}
\email{weisha@zju.edu.cn}
\affiliation{College of Information Science and Electronic Engineering, Zhejiang University, Hangzhou 310027, China}

\author{Xi Dai}\thanks{corresponding author}
\email{daix@ust.hk}
\affiliation{Department of Physics, Hong Kong University of Science and Technology, Clear Water Bay Road, Kowloon, Hong Kong}
\affiliation{Materials Department, University of California, Santa Barbara,
Santa Barbara, CA 93106, USA}

\author{Yue Yu}
\email{yuyue@fudan.edu.cn}
\affiliation{Department of Physics, Fudan University, Shanghai 200438, China}

\date{\today}

\begin{abstract}
This article contains a digest of the theory of electromagnetism and a review of the transformation between inertial frames, especially under low speed limits. The covariant nature of the Maxwell's equations is explained using the conventional language. We show that even under low speed limits, the relativistic effects should not be neglected to get a self-consistent theory of the electromagnetic fields, unless the intrinsic dynamics of these fields has been omitted completely. The quasi-static limits, where the relativistic effects can be partly neglected are also reviewed, to clarify some common misunderstandings and imprecise use of the theory in presence of moving media and other related situations. The discussions presented in this paper provide a clear view of why classical electromagnetic theory is relativistic in its essence. 
\end{abstract}

\maketitle

\section{introduction}

The theory of electromagnetism which unifies various phenomena from electricity and magnetism to light and radio waves is one of the most beautiful, complete, self-consistent and insightful physical theories that have ever been established in human history. Besides being a successful theory of classical physics, it went beyond what had been understood by its creators in the nineteenth century and inspired the revolution of the views on space-time structures in the early twentieth century. The final version of the theory established by James C. Maxwell, together with its relativistic covariant form proposed by Hermann Minkowski, Albert Einstein, Hendrik A. Lorentz, J. Henri Poincaré, et al. is mathematically complete and self-consistent. Without new convincing discoveries from experiments and observations, any attempts to modify or apply the theory without care would result in contradiction \cite{hehl_foundations_2003,zhou_independence_2006}. For instance, relativistic effects occur naturally in electromagnetism, thus neglecting them without caution often leads to irrational results or incorrect predictions to physical measurements. In section \ref{sec_low}, we demonstrate the unsurprising result that no matter how low the relative speed between frames of reference is, relativistic effects in electromagnetism cannot be neglected without extra conditions, which is different with the situation of classical mechanics. 

However, many theories in classical mechanics are constructed based on non-relativistic space-time structure, where the Galilean transformation is used when transforming from one inertial frame to another. Consequently, people are usually only interested in the low speed limits of the electromagnetic theory when investigating classical mechanical systems coupled to electromagnetic fields, provided that all the other speeds $v$ involved are much smaller than the speed of light $c$. A crucial but sometimes forgotten fact is that the accurate transformation between frames as described by H. A. Lorentz does not naturally reduce to Galilean transformation in generic low speed limits ($v/c\ll 1$), but in a non-physical limit $c\rightarrow\infty$ instead. Therefore, one needs to replace Galilean transformation by a correct low speed limit of Lorentz's transformation anywhere in a mechanical theory to make it consistent with electromagnetism, for example, the reconstruction of the acoustic wave equation in presence of moving media done by J. A. Kong \cite{kong_interaction_1970}.

Accurate discussion of the low speed limits also becomes increasingly important in modern condensed matter and material science research when the interactions between matters and electromagnetic fields (such as spectroscopy, photon or X-ray scattering, response to static fields and electromagnetic induction) are receiving more and more attention. In many physical problems, particles usually move at a speed much lower than $c$, constructing the low speed limits of electromagnetism is useful for preserving crucial and non-negligible relativistic effects and keeping the equations simple at the same time. As an example, when matters (media) are moving in the lab frame (``unprimed frame''), the physical laws that electromagnetic fields should obey can be obtained by first writing down the laws in a comoving frame (``primed frame'') where the matters are (locally) stationary and then transforming back to the lab frame \cite{tai_study_1964}. 

In this article, we will mainly focus on the transformation between frames of reference (Lorentz's transformation) for electromagnetism, derive its low speed limits in different situations rigorously and investigate them thoroughly. Lorentz's transformation in a generic situation will be reviewed in Section \ref{lorentz} and used as a starting point for our discussion. After that, we will take the most general form of the low speed limits in Section \ref{sec_low} and show that the theory of electromagnetism remains to be relativistic and form invariant under frame transformation. In Section \ref{quasi}, we show that in addition to the low speed limits, two types of quasi-static limits are often taken to deal with the electromagnetic problems with charge or current densities changing slowly in time. We would emphasize that some of the relativistic effects will be neglected under these quasi-static limits. Next, we give some comments on the inconsistency between electromagnetism and Galilean coordinate transformation and a brief review of the theory of ``Galilean electromagnetism'' known to the engineering community in Section \ref{g_em} from physics perspectives. Finally, a discussion on electromagnetic theory expressed in terms of different quantities and notations is provided in Section \ref{other_forms} for clarification and benefiting research in other areas. The Lagrangian formulation and some discussion on the determination of the theory are provided in Appendix \ref{lagrange}.

\section{transformation between inertial frames}\label{lorentz}

The physical foundation of the low speed expansion approaches to be reviewed in this article is the electromagnetic theory that is consistent with the theory of special relativity \cite{schwinger_classical_1998,jackson_classical_1999}. We will take the theories of electromagnetism and special relativity as the basis to construct a mathematically rigorous framework for applications under low speed limits, especially in condensed matter physics and other relevant fields. To begin with, we briefly review the structure of electromagnetism and the exact form of Lorentz's transformations for coordinates, charge/currents and electromagnetic fields in this section.

The fundamental laws that govern electromagnetic phenomena are expressed as Maxwell's equations
\begin{subequations}\label{maxwell}
\begin{align}
    \vec{\nabla}\cdot\Vec{E} &= \rho/\epsilon_0,\label{columb}\\
    \vec{\nabla}\cdot\Vec{B} &= 0,\label{thompson}\\
    \vec{\nabla}\times\Vec{E} &= -\frac{\partial}{\partial t}\Vec{B},\label{faraday}\\
    \vec{\nabla}\times\Vec{B} &= \mu_0\Vec{J} + \frac{1}{c^2}\frac{\partial}{\partial t}\Vec{E},\label{ampere}
\end{align}
\end{subequations}
where $\Vec{E}$ and $\vec{B}$ are electric and magnetic fields, and $\rho$ and $\vec{J}$ are charge and current densities (usually called charge/currents for simplicity). The above four quantities are all functions of space and time while $\epsilon_0$ and $\mu_0$ are constants which determine the speed of light $c = 1/\sqrt{\epsilon_0\mu_0}$. Equations (\ref{maxwell}) are usually referred to as the ``basic'' form of Maxwell's equations because the charge/currents here include both free and bounded ones. We will stick to this basic form in the following discussions without losing any generality, for the reason that all these four quantities are objective physical quantities without ambiguities in their definitions, which are measurable (directly or indirectly) in experiments. Other forms that involve polarization density $\vec{P}$, magnetization $\vec{M}$, as well as the auxiliary fields: the electric displacement field $\vec{D}$ and ``the H-field'' (magnetic field strength) $\vec{H}$ will be discussed afterward in Section \ref{other_forms}.

It is said that electric field $\vec{E}$ and magnetic field $\vec{B}$ are objective and measurable quantities because they are the direct origins of electromagnetic forces (Lorentz's forces) $\vec{F}$ on a moving (point) particle with charge $q$ and velocity $\vec{u}$, which can be measured by inspecting the motion of the particle $d\vec{p}/dt$:
\begin{equation}\label{dynamics}
    \vec{F} = q(\vec{E} + \vec{u}\times\vec{B}) = \frac{d\vec{p}}{dt}.
\end{equation}
This serves as the foundation for the further discussion of transformations of $\vec{E}$ and $\vec{B}$ between frames of reference due to the fact that the correct transformation should provide the correct values of measurable quantities in one frame in terms of a same set of quantities in another frame, or in another word, to make (\ref{dynamics}) have the same form in any frame. 

The construction of the whole theory of frame transformation is presented in many text books, where Chapter 12 of \textit{Electromagnetism} by Gerald L. Pollack and Daniel R. Stump\cite{pollack_electromagnetism_2002} is one of the best. The most important results are the transformation of coordinates $(\vec{r},t)$:
\begin{subequations}\label{coord_trans}
\begin{align}
    &\vec{r}'_\parallel = \gamma(\vec{r}_\parallel-\vec{v}t),\\
    &\vec{r'}_\perp = \vec{r}_\perp,\\
    &t' = \gamma(t-\vec{v}\cdot\vec{r}/c^2),
\end{align}
\end{subequations}
the transformation of charge/currents $(\vec{J},\rho)$:
\begin{subequations}\label{current_trans}
\begin{align}
    &\vec{J}'_\parallel = \gamma(\vec{J}_\parallel-\vec{v}\rho),\\
    &\vec{J'}_\perp = \vec{J}_\perp,\\
    &\rho' = \gamma(\rho-\vec{v}\cdot\vec{J}/c^2),
\end{align}
\end{subequations}
and the transformation of electromagnetic fields:
\begin{subequations}\label{field_trans}
\begin{align}
    &\vec{E'}_\perp = \gamma(\vec{E}_\perp+\vec{v}\times\vec{B}_\perp),\\
    &\vec{E}'_\parallel = \vec{E}_\parallel,\\
    &\vec{B'}_\perp = \gamma(\vec{B}_\perp-\vec{v}\times\vec{E}_\perp/c^2),\\
    &\vec{B}'_\parallel = \vec{B}_\parallel.
\end{align}
\end{subequations}
In the above transformations, $\vec{v}$ is the relative velocity of the primed frame measured in the unprimed frame, while the parallel $\parallel$ or perpendicular $\perp$ is with respect to it. The paramter $\gamma = 1/\sqrt{1-v^2/c^2}$ is the Lorentz factor.  The inverse transformations are obtained by interchanging the primed and unprimed quantities and replacing $\vec{v}$ with $-\vec{v}$, or by solving the unprimed quantities from the above transformation equations in terms of primed quantities. Since $(\vec{r},t)$ and $(\vec{J},\rho)$ transform according to Lorentz's transformation as the 4-dimensional vectors in the Minkowsky space, while the components of $\vec{E}$ and $\vec{B}$ form a tensor which also transforms according to Lorentz's transformation (see Appendix \ref{lagrange} (\ref{tenser_trans})), they are said to be ``covariant'' under frame transformation. 

The three sets of transformations (\ref{coord_trans}), (\ref{current_trans}) and (\ref{field_trans}) can be seen as the results of three distinct experiments measuring coordinates, charge/currents and electromagnetic fields in both frames. If the kinematics of the charge-carrying particles is included in the theory, (\ref{current_trans}) can be derived from (\ref{coord_trans}) by applying the transformation of the coordinates so that there is no need to perform an extra experiment to verify (\ref{current_trans}), while (\ref{field_trans}) is totally independent of them, which can be verified by a separate set of experimental measurements.

To investigate the transformation of the local relations of the four quantities, i.e. Maxwell's equations, between different frames, we first write down the equations (\ref{maxwell}) in a comoving (primed) frame (remember to add primes to everything in (\ref{maxwell}), such as $\nabla\rightarrow\nabla'$ and $\partial/\partial t\rightarrow\partial/\partial t'$), and then transform everything to the lab (unprimed) frame in which the comoving frame is moving at a velocity $\vec{v}$ by making substitutions (\ref{coord_trans}), (\ref{current_trans}) and (\ref{field_trans}). Here we would like to emphasize that since Maxwell's equations are partial differential equations, properly transforming the differential operators is needed:
\begin{subequations}\label{d_trans}
\begin{align}
    \vec{\nabla}'_{\parallel} &= \gamma(\vec{\nabla}_{\parallel} + \frac{\vec{v}}{c^2}\frac{\partial}{\partial t}),\\
    \vec{\nabla}'_{\perp} &= \vec{\nabla}_{\perp},\\
    \frac{\partial}{\partial t'} &= \gamma(\frac{\partial}{\partial t} + \vec{v}\cdot\vec{\nabla}),
\end{align}
\end{subequations}
and inversely:
\begin{subequations}
\begin{align}
    \vec{\nabla}_{\parallel} &= \gamma(\vec{\nabla}'_{\parallel} - \frac{\vec{v}}{c^2}\frac{\partial}{\partial t'}),\\
    \vec{\nabla}_{\perp} &= \vec{\nabla}'_{\perp},\\
    \frac{\partial}{\partial t} &= \gamma(\frac{\partial}{\partial t'} - \vec{v}\cdot\vec{\nabla}').
\end{align}
\end{subequations}

After these substitutions, we will find that Maxwell's equations (\ref{maxwell}) is form invariant under frame transformations, see \cite{jefimenko_relativistic_1999} as an example. 

Here we would like to revisit the principle of relativity. In Einstein's famous paper on the electrodynamics of moving bodies \cite{einstein_zur_2005}, the principle was stated as:
\begin{quote}
``The laws by which the states of physical systems undergo change are not affected, whether these changes of state be referred to the one or the other of
two systems of co-ordinates in uniform translatory motion. ''
\end{quote}
One should not confuse the physical law itself with its mathematical representation, which is usually written as a (partial) differential equation. The physical law is not affected under frame transformation, as the principle of relativity states, does not directly imply that the differential equation stays unchanged. The covariance of Maxwell's equations is grounded by the principle of relativity, together with the fact that electromagnetic fields are matters themselves which are independent of any media, which guarantees that there is no "privileged" frame of reference. These two points combined together ensure that the physical laws describing the electromagnetic phenomena can be written down in terms of covariant 4-vectors and tensors which transform according to Lorentz's transformation, thus the final equations are form invariant under frame transformation. As a counterexample, sound waves are not independent matter but the disturbance of material media \cite{ferraro_einsteins_2007}, the corresponding dynamic equations of sound waves then must involve terms that can not be written as the covariant 4-vectors and their forms will be changed under frame transformations \cite{kong_interaction_1970}.

In fact, this was an old problem that Maxwell recognized when he built up his equations. He found his equations were not invariant under (Galilean) frame transformations and valid only in a unique ``Ether frame'' (J. C. Maxwell, \textit{The Encyclopedia Britannica}, 9th ed. 1875-1889; reprinted in The Scientific Papers of James Clark Maxwell, ed. by W. D. Niven), see Section \ref{g_em} for more detailed discussions. The theory of special relativity gives the correct transformation between frames and makes Maxwell's equations have the same form in any (inertial) frame of reference.

\section{The general form of low speed limits}\label{sec_low}

Having understood how to do frame transformations generically in presence of electromagnetic fields and sources (charge/current densities), we continue introducing the low speed limits of it because $v/c\ll 1$ is usually the case in today's research of condensed matter physics and material science, and preserving the exact form of equations (\ref{coord_trans}), (\ref{current_trans}) and (\ref{field_trans}) in such a situation is sometimes unnecessary. By saying the ``general'' form of low speed limits we mean taking the first few terms in the series expansion of Lorentz's transformation with respect to the dimensionless parameter $\beta=v/c$, without any other approximation. Other ``special'' forms with extra approximations will be discussed in Section \ref{quasi}.

There are actually two different ways of establishing the low speed limit. One is to take the full version of the theory mentioned in Section \ref{lorentz} as the starting point, and then take only the first few terms in series expansion; the other way is to take an approximate form of only the coordinate transformation, then establish the transformation of charge/currents based on the kinematics of charged particles, and finally follow the same method in Einstein's paper \cite{einstein_zur_2005} to look for the right transformation of the fields that makes Maxwell's equations form invariant. In this section, we will follow the first route because much caution is needed to get mathematically correct and physically meaningful results by the second method.

Before doing series expansion, it is beneficial to multiply a $c$ on both sides of the last equation in each of the above transformations (\ref{coord_trans}) and (\ref{current_trans}), and take $ct$ and $c\rho$ to be the quantities we considerate, because they have the same dimensions as $\vec{r}$ and $\vec{J}$, respectively. The same operation should also be done on the last two equations in (\ref{field_trans}), so that $c\vec{B}$ and $\vec{E}$ have the same dimension. Now, we see that each $v$ in any of the above transformations appears together with a $c$ as a dimensionless parameter $\beta = v/c$. 

In particular, we preserve up to the first order terms in the expansions, which leads to
\begin{subequations}\label{coord_low}
\begin{align}
    &\vec{r'} = \vec{r}-\vec{v}t,\\
    & t' = t-\frac{\vec{v}\cdot\vec{r}}{c^2},
\end{align}
\end{subequations}
\begin{subequations}\label{current_low}
\begin{align}
    &\vec{J'} = \vec{J}-\vec{v}\rho,\\
    & \rho' = \rho-\frac{\vec{v}\cdot\vec{J}}{c^2},
\end{align}
\end{subequations}
\begin{subequations}\label{field_low}
\begin{align}
    &\vec{E'} = \vec{E}+\vec{v}\times\vec{B},\\
    &\vec{B'} = \vec{B}-\frac{\vec{v}\times\vec{E}}{c^2},
\end{align}
\end{subequations}
and
\begin{subequations}\label{d_low}
\begin{align}
    \vec{\nabla}' &= \vec{\nabla} + \frac{\vec{v}}{c^2}\frac{\partial}{\partial t},\\
    \frac{\partial}{\partial t'} &= \frac{\partial}{\partial t} + \vec{v}\cdot\vec{\nabla},
\end{align}
\end{subequations}

From the above equations (\ref{coord_low}), (\ref{current_low}) and (\ref{d_low}) we can clearly see that even in low speed limits, Lorentz's transformation does not reduce to the following Galilean transformation
\begin{subequations}\label{coord_g}
\begin{align}
    \vec{r}' &= \vec{r} - \vec{v}t,\\
    t' &= t,
\end{align}
\end{subequations}
\begin{subequations}\label{current_g}
\begin{align}
    &\vec{J'} = \vec{J}-\vec{v}\rho,\\
    & \rho' = \rho,
\end{align}
\end{subequations}
and
\begin{subequations}\label{d_g}
\begin{align}
    \vec{\nabla}' &= \vec{\nabla},\\
    \frac{\partial}{\partial t'} &= \frac{\partial}{\partial t} + \vec{v}\cdot\vec{\nabla}
\end{align}
\end{subequations}
no matter up to which order is preserved. If we preserve only the zeroth order terms, what we get is the "stationary limit" that there is no relative motion between two frames at all. Up to the first order, it is fundamentally different from Galilean transformation. As a result, it is unsuitable to say that the low speed limit is ``non-relativistic''. The actual and mathematically rigorous non-relativistic limit is $c\rightarrow\infty$, which is a non-physical limit\cite{kong_interaction_1970}. 

With most of the relativistic kinetic effects such as time dilation and length contraction being neglected (because we take $\gamma\approx 1$), the essential difference between the first order low speed limit of the Lorentz's coordinate transformation (\ref{coord_low}) and the Galilean coordinate transformation (\ref{coord_g}) is in their time components, where the equation (\ref{coord_low}) indicates a temporal shift that is symmetric to the spatial shift. This shift in time reflects the fact that the relativity of simultaneity is a first order effect that should not be neglected even under the low speed limit. 

Before proceeding to apply these transformations, we would like to emphasize that they are the first order approximations, any higher order terms appeared in the calculations involving these transformations should be omitted to keep the approximation self-consistent. Additionally, the inverse of these transformations must be obtained by exchanging the primed and unprimed quantities and reversing $\vec{v}$. (If you want to solve for unprimed quantities in terms of primed quantities, you will need to discard the higher order terms of $v$.)

With every piece in (\ref{coord_low}), (\ref{current_low}), (\ref{field_low}) and (\ref{d_low}) being prepared, we can now imitate the procedure in \cite{jefimenko_relativistic_1999} to see how physics laws change under the frame transformation in the low speed limit. Actually, before diving into massive calculations, we have a firm belief that Maxwell's equations will not change their forms under transformations under the low speed limit, at least up to the proper order. The underlying reason is that since Maxwell's equations are form invariant under the full Lorentz's transformation, they should also be form invariant up to the corresponding order we keep. The results in Appendix \ref{append_trans} indeed prove our belief. Moreover, the speed of light will also stay the same in any arbitrary frame under the low speed limit, and an approximate form of relativistic Doppler's effect can also be obtained under the low speed limit, see Appendix \ref{doppler} for details. 

As a summary, we emphasize that the relativistic natures of electromagnetism, such as covariance, unchanged speed of light in any frame and relativistic Doppler effect, are still preserved in the general form of the low speed limits. The theory of electromagnetism is essentially relativistic and expressed precisely by mathematics, thus to get the correct theoretical results for electromagnetic phenomena, the effects of relativity should not be neglected no matter how low the speed $v$ is.

\section{The quasi-static limit}\label{quasi}
In this section, we discuss some special physical situations, in which some of the relativistic effects can be neglected. First, we know that relativity does not show up in the electrostatics and magnetostatics because fields $\vec{E}$ and $\vec{B}$ are determined by their corresponding sources $\rho$ and $\vec{J}$ instantly. Under such a static limit, Maxwell's equations look like
\begin{subequations}\label{static}
\begin{align}
    \vec{\nabla}\cdot\Vec{E} &= \rho/\epsilon_0,\\
    \vec{\nabla}\cdot\Vec{B} &= 0,\\
    \vec{\nabla}\times\Vec{E} &= 0,\\
    \vec{\nabla}\times\Vec{B} &= \mu_0\Vec{J}.
\end{align}
\end{subequations}
All the time derivatives are gone due to the static condition. However, the static limit is much trickier than it appears to be: on one hand, under such a limit the electromagnetic fields $\vec{E}$ or $\vec{B}$ is no longer a special type of independent matter by themselves that can be replaced by some types of instant interaction between charge or current; on the other hand, the frame transformation is not allowed here. It can be seen that after transforming to another frame, we will generically not be in the static limit anymore. 

A solution to this dilemma is to divide the static limit into two different limits: electrostatic limit and magnetostatic limit, in which only one of the two fields/sources dominates. After such a partition, we can weaken the static condition to be ``quasi-static'' to allow the fields and sources to vary slowly in time. Since the quasi-static limit is in addition to the general low speed limit, every time-varying term comes from frame transformation will be in the order of $v/c$ compared with the original terms, which keeps us still in the quasi-static limit. 

Similar to the condition of low speed limits that the ratio $v/c$ is a small (dimensionless) number ($v/c\ll 1$), the condition of the additional quasi-static limits is that for any physical quantity $G$ involved, the ratio
\begin{equation}
    \frac{L}{cG}\frac{\partial G}{\partial t}
\end{equation}
is a small number with the same order of $v/c$, where $L$ is a typical spatial (length) scale of the system being studied.

For the electric quasi-static limit, we take the electric fields and charge to be dominant, while ratios $cB/E$ and $J/c\rho$ are both with the same order of $v/c$. By keeping the two leading (zeroth and first) orders in (\ref{maxwell}), (\ref{current_low}) and (\ref{field_low}), together with the first order terms in $cB/E$ and $J/c\rho$ as indicated above, we get the following field equations in the electric limit of the quasi-static approximation:
\begin{subequations}\label{electriceq}
\begin{align}
    \vec{\nabla}\cdot\Vec{E} &= \rho/\epsilon_0,\\
    \vec{\nabla}\cdot\Vec{B} &= 0,\\
    \vec{\nabla}\times\Vec{E} &= 0,\label{ele3}\\
    \vec{\nabla}\times\Vec{B} &= \mu_0\Vec{J} + \frac{1}{c^2}\frac{\partial}{\partial t}\Vec{E},
\end{align}
\end{subequations}
the transformation of charge/currents:
\begin{subequations}\label{current_e}
\begin{align}
    &\vec{J'} = \vec{J}-\vec{v}\rho,\\
    & \rho' = \rho,
\end{align}
\end{subequations}
and the transformation of the fields between different frames:
\begin{subequations}\label{field_e}
\begin{align}
    &\vec{E'} = \vec{E},\\
    &\vec{B'} = \vec{B}-\frac{\vec{v}\times\vec{E}}{c^2}.
\end{align}
\end{subequations}
For the electric limit, we see that the magnetic field is decoupled from the electric field. The divergence-free (transverse) part of $\vec{E}$ is neglected while both $\vec{B}$ and curl-free (longitudinal) part of $\vec{E}$ are determined instantly by the corresponding sources (both $\vec{J}$ and $\partial\vec{E}/\partial t$ are regarded as the sources of $\vec{B}$ here). A practical example in modern physics research of this limit is the plasmon modes in solid metals \cite{pines_collective_1952}.

Similarly, for the magnetic quasi-static limit, we take magnetic fields and currents to be dominant, while ratios $E/cB$ and $c\rho/J$ are both as the same order of $v/c$. Again, by keeping only the two leading orders in (\ref{maxwell}), (\ref{current_low}) and (\ref{field_low}), we get the field equations under the magnetic quasi-static limit as
\begin{subequations}\label{magneticeq}
\begin{align}
    \vec{\nabla}\cdot\Vec{E} &= \rho/\epsilon_0,\\
    \vec{\nabla}\cdot\Vec{B} &= 0,\\
    \vec{\nabla}\times\Vec{E} &= -\frac{\partial }{\partial t}\vec{B},\\
    \vec{\nabla}\times\Vec{B} &= \mu_0\Vec{J},\label{mag4}
\end{align}
\end{subequations}
the transformation of charge/currents:
\begin{subequations}\label{current_m}
\begin{align}
    &\vec{J'} = \vec{J},\\
    & \rho' = \rho-\frac{\vec{v}\cdot\vec{J}}{c^2},
\end{align}
\end{subequations}
and transformation of fields:
\begin{subequations}\label{field_m}
\begin{align}
    &\vec{E'} = \vec{E}+\vec{v}\times\vec{B},\\
    &\vec{B'} = \vec{B}.
\end{align}
\end{subequations}
It is clear that the longitudinal electric field is decoupled from the magnetic field and both transverse $\vec{E}$ and $\vec{B}$ are determined by their sources instantly. A practical example of this limit is the studies of vortex motion in type-II superconductors \cite{landau_theory_1965,jiang_quantum_2019}. In both of the above two limits, the form invariant nature of field equations is preserved.

We would like to emphasize here that no further approximation has been made about the coordinates, thus their transformation should still be (\ref{coord_low}) instead of something else. In Section \ref{lorentz} we mentioned that when including the kinematics of the charge-carrying particles (the microscopic origin of the charge and current densities) into the theory, transformation of charge/currents can be derived merely from the transformation of coordinates. The reason why a same coordinate transformation (\ref{coord_low}) can generate different transformations of charge/currents is that we neglected higher order terms of $J/c\rho$ or $c\rho/J$ respectively in (\ref{current_e}) and (\ref{current_m}).

The final question of this section is: Are the relativistic effects negligible under these quasi-static limits? Intuitively we might want to check whether the covariance of the theory is still preserved if coordinates transform according to Galilean transformation (\ref{coord_g}), instead of the low speed limit of Lorentz transformation (\ref{coord_low}). This was studied in the 1970s by M. Le Bellac, J.M. Levy-Leblond, G. Rousseaux, et al., confirming that if one uses Galilean coordinates transformation in these two quasi-static limits, the forms of the corresponding equations do not change. More details on their work are discussed in the next section.

Although Galilean coordinates transformation (\ref{coord_g}) preserves the covariance of the theory in both limits,  we find that only under electric quasi-static limit, charge/currents transformation (\ref{current_e}) is consistent with the coordinate transformation (\ref{coord_g}); while the same consistency can not be reached for the magnetic quasi-static limit once we consider the microscopic origin of the electromagnetic sources (charge and current densities).  

Consequently, we claim that electrostatic effects are somewhat alike classical mechanical phenomena so that relativistic effects can be safely neglected in electric quasi-static limit, while magnetostatic effects are not. As Richard P. Feynman once said in his famous lectures on physics that ``magnetism is in reality a relativistic effect of electricity'' \cite{feynman_feynman_2006}, one should never omit the seemingly small relativistic effect because it is the only effect appears when magnetic fields are dominant. The physical reason why in the electric quasi-static limit we can safely neglect the relativistic effects is that the electromagnetic fields in such a limit lose their own dynamics completely and are not independent matters anymore. Once again, we see relativity is truly the essence of electromagnetism.

\section{Electromagnetism and Galilean transformation}\label{g_em}

Recently, there has been a controversy about how Maxwell's equations transform under Galilean transformation \cite{sheng_lorentz_2022}. The Galilean transformation only has space-time part, which means it only transforms the coordinates and charge/currents but never recombines the components of fields as in (\ref{field_low}). The underlying physical meaning is that all the local physical quantities, including the electromagnetic fields, are all "attached" to the matter distributed over the space, which is called medium and can in general be moving. This is the origin of the concept ``Ether'', which was imagined to exist as the medium of electromagnetic fields in history. The ``local'' motion of the medium should be discussed in a comoving frame that it is not moving in a stream, or properly transform all the coordinates to another frame according to coordinate transformation. In a recent paper, its author actually did this to Maxwell's equations in the comoving frame, and obtained some ``material derivative'' terms $(\partial/\partial t + \vec{v}\cdot\vec{\nabla})$ in the equations back to the lab frame, replacing the correct $\partial/\partial t$ \cite{wang_expanded_2021}. Their procedure followed the historical perspective that electromagnetic waves are the local motion of ``Ether''. Measurable consequences that can be inferred from the ``expanded'' Maxwell's equations, such as the frame dependence of the speed of light and the space inhomogeneity of the magnetic field creates the electric field, were never observed in any experiments.

From the modern perspectives, it is meaningless to discuss whether Maxwell's equations change their forms under Galilean transformation, because we know that the electromagnetic fields are not carried by any local motion of some special (baryonic) matter, but a certain kind of matter themselves, which should be transformed properly between frames (similar to the transformation of charge/currents). We can of course set up a strategy of transforming the components of electromagnetic fields between frames in order to preserve the form of Maxwell's equations when coordinates are transformed by Galilean transformation, but this theory will no longer have physical meanings because the made-up transformation might not provide correct $\vec{E}$ and $\vec{B}$ fields as well as $\rho$ and $\vec{J}$ that are measured in another frame. This is to say, electromagnetism is inconsistent with Galilean transformation because electromagnetism, especially magnetism is relativistic in its essence. 

The only strict way of making electromagnetism consistent with the Galilean transformation is to take the non-physical limit $c\rightarrow\infty$, though it is meaningless because we will run into a dilemma of not knowing the physical interpretation of the theory. However, it reminds us that in some special situations the electromagnetic theory can possibly be consistent with Galilean transformation, as long as the light travels fast enough that time retardation caused by it can be neglected in the problem of interest. 

This topic has been studied since the 1970s, which formed an area named ``Galilean electromagnetism'' \cite{rousseaux_forty_2013,montigny_electrodynamics_2006,de_montigny_applications_2007} that is well known in areas like electrical and mechanical engineering, though not many physicists have heard of it. We emphasize here that ``Galilean electromagnetism'' is not an alternative to special relativity but is derived from the low speed limits in some special situations. See \cite{le_bellac_galilean_1973} for more information. 

From our point of view, the situations considered by ``Galilean electromagnetism'', called ``electric limit'' and ``magnetic limit'', are just the two quasi-static limits we discussed in Section \ref{quasi}, but replacing coordinates transformation (\ref{coord_low}) with Galilean coordinates transformation (\ref{coord_g}). There is no problem for the electric limit, but a contradiction is seen in the magnetic limit because the desired ways to transform charge/current densities (\ref{current_m}) can not be self consistently derived from the transformation of the coordinates listed in (\ref{coord_g}), if we consider the microscopic origin of the sources. This problem is due to the logic of ``Galilean electromagnetism''. It is to first require the invariant nature of field equations under Galilean coordinates transformation, and then look for other pieces of the theory (transformations of charge/currents and fields) to satisfy that requirement, which might not be consistent with the transformation of the coordinates. 

Therefore, our conclusion is that only the ``Galilean electromagnetism'' under the electric limit is consistent with the Galilean transformation. This is because the intrinsic dynamics of the electric and magnetic fields has been completely  neglected and the fields in such a quasi-static approximation can be determined instantly by the charge/current densities only, which makes the fields instantly follow the motion of the sources and not surprisingly lose their relativistic nature.

\section{Discussion on other forms of Maxwell's equations}\label{other_forms}

In the previous sections of this article, we restrict our discussion to the basic form of Maxwell's equations which only involves the measurable quantities $\vec{E}$, $\vec{B}$, $\rho$ and $\vec{J}$ to avoid ambiguity. In principle, we can describe any classical electromagnetic problem with these four quantities, but in the presence of media, people conventionally separate the bounded charge/currents: $(\rho_{\rm b},\vec{J}_{\rm b})$ induced in media from the free charge/currents: $(\rho_{\rm f},\vec{J}_{\rm f})$, that is
\begin{subequations}
\begin{align}
    \rho = \rho_{\rm f} + \rho_{\rm b}\\
    \vec{J} = \vec{J}_{\rm f} + \vec{J}_{\rm_b}.
\end{align}
\end{subequations}
Both free and bounded charge/currents are measurable quantities. In order not to be troubled with microscopic variations of the quantities which result from the molecular structure of matter, we follow the convention of interpreting every quantity involved to be its averaged value over elements of volume which are ``physically infinitesimal'' from a macroscopic perspective. By taking this average, we hide the microscopic structure of the matter as well as the fast spatial oscillation of the fields at the atomic scale, which brings us to the ``continuous media limit'' \cite{landau_electrodynamics_2009}.

In order to provide more detailed descriptions of the response of the media to electromagnetic fields, two extra quantities: electric polarization density $\vec{P}$ and magnetization $\vec{M}$ are introduced respectively as the sum of electric and magnetic dipole moments in a unit volume, averaging over a macroscopic but still small enough volume comparing to the scale of interest, just like for other quantities. Their relations with $(\rho_{\rm b},\vec{J}_{\rm b})$ are
\begin{subequations}\label{current_b}
\begin{align}
    \rho_{\rm b} &= -\vec{\nabla}\cdot\vec{P},\\
    \vec{J}_{\rm b} &= \frac{\partial\vec{P}}{\partial t} + \vec{\nabla}\times\vec{M}.
\end{align}
\end{subequations}
It is worth mentioning that these two equations do not uniquely define $\vec{P}$ and $\vec{M}$. Polarization and magnetization are determined by their definitions above only if the microscopic structure of the media is taken into account (for the simplest example, adding the response of $\vec{P}(\vec{E})$ to $\vec{E}$ manually). 

Our next question is about the frame transformations of $\vec{P}$ and $\vec{M}$. Just like for charge/currents, their transformations can be found by the transformation of coordinates (\ref{coord_trans}) and the kinematics of the charged particles in media. Please refer to the corresponding chapters in Pauli's lecture on electrodynamics \cite{pauli_electrodynamics_2000}. The final results can be summarised as
\begin{subequations}\label{pm_trans}
\begin{align}
    &\vec{M}'_\perp = \gamma(\vec{M}_\perp+\vec{v}\times\vec{P}_\perp),\\
    &\vec{M}'_\parallel = \vec{M}_\parallel,\\
    &\vec{P}'_\perp = \gamma(\vec{P}_\perp-\vec{v}\times\vec{M}_\perp/c^2),\\
    &\vec{P}'_\parallel = \vec{P}_\parallel.
\end{align}
\end{subequations}
The correct transformations of measurable quantities $\rho_{\rm b}$ and $\vec{J}_{\rm b}$ are the same as the free ones (\ref{current_trans})), as one can check by connecting (\ref{pm_trans}) and (\ref{current_b}). The similarity in (\ref{pm_trans}) and (\ref{field_trans}) indicates that components of $\vec{P}$ and $\vec{M}$ can also have a covariance form.

The most popular form of Maxwell's equations in presence of media is the Minkowski's form which includes ``auxiliary fields'' defined by
\begin{subequations}
\begin{align}
    \vec{D} &= \epsilon_0\vec{E} + \vec{P},\\
    \vec{H} &= \frac{\vec{B}}{\mu_0} - \vec{M}.
\end{align}
\end{subequations}
The Minkowski's form of the equations is
\begin{subequations}\label{minkowski}
\begin{align}
    \vec{\nabla}\cdot\Vec{D} &= \rho_{\rm f},\label{columb_m}\\
    \vec{\nabla}\cdot\Vec{B} &= 0,\label{thompson_m}\\
    \vec{\nabla}\times\Vec{E} &= -\frac{\partial}{\partial t}\Vec{B},\label{faraday_m}\\
    \vec{\nabla}\times\Vec{H} &= \Vec{J}_{\rm f} + \frac{\partial}{\partial t}\Vec{D},\label{ampere_m}
\end{align}
\end{subequations}
Equations (\ref{minkowski}) are covariant (form invariant under frame transformation) because the covariance of both $\vec{E}$, $\vec{B}$ and $\vec{P}$, $\vec{M}$ determines that $\vec{D}$ and $\vec{H}$ also transform covariantly between different frames.  

The above discussion about the transformation of $\vec{P}$ and $\vec{M}$ and the form invariance of Minkowski's form can be directly applied to low speed limits by taking $\gamma = 1$ for the first order. Here we would like the readers to realize that when introducing new quantities in mathematical formulas, attentions are needed to ensure everything described by mathematical symbols has the exact connections to measurable physical quantities. Measurable quantities like $\vec{E}$, $\vec{B}$, $\rho$ and $\vec{J}$ cannot be arbitrarily defined, so as their transformations between frames. In a classical paper published in 1976, the authors obtained the correct result of form invariance of Maxwell's equations under deformation by introducing a set of transformations of the fields that meet the requirement, which turned out to mix different orders in the expansion. \cite{lax_maxwell_1976}.

Additionally, we have to emphasize that although the form of equations is invariant under frame transformation, the constitutive relations do change. For example, if the relations in the comoving (primed) frame are:
\begin{align*}
    \vec{D}' &= \epsilon\vec{E}',\\
    \vec{B}' &= \mu\vec{H}',\\
    \vec{J}' &= \sigma\vec{E}',
\end{align*}
the set of relations in the lab (unprimed) frame would be \cite{tai_study_1964}:
\begin{align*}
    \vec{D} &= \epsilon\vec{E} + (\epsilon\mu-\epsilon_0\mu_0)\vec{v}\times\vec{H},\\
    \vec{B} &= \mu\vec{H} - (\epsilon\mu-\epsilon_0\mu_0)\vec{v}\times\vec{E},\\
    \vec{J}_{\rm f} &= \rho_{\rm_f}\vec{v} + \sigma(\vec{E} + \vec{v}\times\vec{B}),
\end{align*}
if we only preserve up to the first order of $v/c$. (Here $\rho_{\rm f}$ is the free charge distributed in the media measured in the lab frame. Remember $\rho_{\rm f}'\vec{v} = \rho_{\rm f}\vec{v}$ when preserving up to first order.)

Another well-known form of Maxwell's equations was developed by L. J. Chu \cite{chu_electromagnetic_1960}. We find it a bit misleading because Chu used a set of quantities (with subscript $c$) that are linear combinations of the usual physical quantities to write down his equations:
\begin{align*}
    \vec{E}_c + \mu_0\vec{M}_c\times\vec{v} = \vec{E},\\
    \mu_0(\vec{H}_c + \vec{M}_c) = \vec{B},\\
    \vec{H}_c - \vec{P}_c\times\vec{v} = \vec{H},\\
    \epsilon_0\vec{E}_c + \vec{P}_c = \vec{D},\\
    \vec{J}_c = \vec{J}_{\rm f},\\
    \rho_c = \rho_{\rm f}.
\end{align*}
The equations are identical to Minkovski's form:
\begin{subequations}\label{chu}
\begin{align}
    &\vec{\nabla}\cdot(\epsilon_0\vec{E}_c + \vec{P}_c) = \rho_c,\label{columb_c}\\
    &\vec{\nabla}\cdot\mu_0(\vec{H}_c + \vec{M}_c) = 0,\label{thompson_c}\\
    &\vec{\nabla}\times(\vec{E}_c + \mu_0\vec{M}_c\times\vec{v}) = -\frac{\partial}{\partial t}\mu_0(\vec{H}_c + \vec{M}_c),\label{faraday_c}\\
    &\vec{\nabla}\times(\vec{H}_c - \vec{P}_c\times\vec{v}) = \Vec{J}_c + \frac{\partial}{\partial t}(\epsilon_0\vec{E}_c + \vec{P}_c),\label{ampere_c}
\end{align}
\end{subequations}
but most of the quantities with subscript $c$ does not have the same physical meanings as they appear to be. The only advantage of Chu's form is that the constitutive relations in presence of moving media have simpler form, see \cite{tai_study_1964}. Chu's form is not invariant under frame transformation as well.

In summary, though involving the auxiliary fields, Minkowski's form of Maxwell's equations are both form invariant under frame transformation and gauge invariant. In presence of moving media, the constitutive relations for Minkowski's form of equations change their forms under frame transformation.

\section{conclusion}
In conclusion, we have provided a thorough review of the frame transformation in electromagnetism, and show that in general low speed limits relativistic effects cannot be neglected, unless a quasi-static limit is added to it. Maxwell's equations always remain form invariant under transformations in all the limits we reviewed, due to the reason that it is invariant under the complete relativistic transformation. Attempt to establish a self-consistent theory of electromagnetism under Galilean transformation is physically meaningless. The motions of media do not change the form of Maxwell's equations but the constitutive relations. Most importantly, the electromagnetic theory is relativistic in its essence, due to the fact that magnetism is really a relativistic effect of electricity.

We would like to emphasize once again that any discussion in theoretical physics should be grounded in experimental observables, which appear as measurable quantities in theories. Any attempt to establish or modify a theory without confirming whether the new theory yields correct results for measurable quantities will end up with contradictions. The theory of electrodynamics is classical but not outdated. A hundred years after its completion, thinking about details of its physical foundations and mathematical structures can still enhance our understanding of nature and inspire both modern physics and engineering research.

\section*{acknowledgement}
The authors are grateful to all the people, have or haven't met in person, who discussed online about many relevant topics in classical electromagnetism, especially those who have the courage to express their opinions on the internet no matter they turn out to be correct or not in the end. Specifically, the insightful conversations with Meng Cheng, Congjun Wu, Xiaofeng Jin, Jian Zi, Minxing Luo, Weishi Wan, Hakan T{\"u}reci, Alejandro Rodriguez, Duncan Haldane, Chenyu Zhu, Xueqi Chen, Qingyi Zhou, Junlan Jin, Yiyi Li and Thomas Maldonado greatly helped in clarifying misunderstanding of detailed concepts. The authors are also thankful for the comments from Alisia Pan, Cheng Wen and Xinyu Zhang which help improve this paper. The authors acknowledge the moral support and suggestions from Biao Lian, Abhinav Prem, Zhi Liu, Xuguang Xu, Fuchun Zhang and Changpu Sun as well as many other people from Princeton University, ShanghaiTech University, Institute of Physics and other institutions under the Chinese Academy of Sciences.

\onecolumngrid
\appendix
\section{Lagrangian formulation of classical electromagnetism }\label{lagrange}

\subsection{Maxwell equations}

It is known that in field theory Maxwell's equations can be written compactly in terms of electromagnetic tensor
\begin{equation}
    F_{\mu\nu} = \partial_{\mu}A_{\nu} - \partial_{\nu}A_{\mu},
\end{equation}
where $A_{\mu}$ is a component of the 4-potential. We use Gaussian units with $c=1$. The tensor $F_{\mu\nu}$ is an anti-symmetric covariant tensor which transform between frames as
\begin{equation}\label{tenser_trans}
    F'^{\mu\nu}(x') = \Lambda^{\mu}_{\;\;\rho}\Lambda^{\nu}_{\;\;\sigma}F^{\rho\sigma}(x),
\end{equation}
where $x$ represents $(t,x,y,z)$ is the space-time coordinate.

Following the orthodox route, we first write down the Lagrangian density of an electrodynamics system:
\begin{equation}
    \mathcal{L} =  - \frac{1}{4}F_{\mu\nu}F^{\mu\nu}-J^\mu A_{\mu}+\mathcal{L}_{matter},
\end{equation}
where $J^\mu$ is the 4-current of the matter field and $\mathcal{L}_{matter}$ is the Lagrangian density of the matter field.   From here and now on, we use Einstein summation convention for neat notation.

According to  the Euler-Lagrange equation for the electromagnetic 4-potential $A_\nu$, 
\begin{equation}
    \partial_{\mu}\left(\frac{\partial\mathcal{L}}{\partial(\partial_{\mu}A_\nu)}\right) - \frac{\partial\mathcal{L}}{\partial A_\nu} = 0,
\end{equation}
we compute the following quantities:
\begin{equation}
    \frac{\partial\mathcal{L}}{\partial A_\nu} = -J^{\nu}
\end{equation}
 and
\begin{align*}
    &\left(\frac{\partial\mathcal{L}}{\partial(\partial_{\mu}A_\nu)}\right) =-\frac{1}{4}\frac{\partial(F_{\alpha\beta}F^{\alpha\beta})}{\partial(\partial_\mu A_\nu)}=-\frac{1}{4}g^{\alpha\lambda}g^{\beta\sigma}\frac{\partial(F_{\alpha\beta}F_{\lambda\sigma})}{\partial(\partial_\mu A_\nu)}\\
    &=-\frac{1}{4}g^{\alpha\lambda}g^{\beta\sigma}\left[\frac{\partial(F_{\alpha\beta})}{\partial(\partial_\mu A_\nu)}F_{\lambda\sigma} + \frac{\partial(F_{\lambda\sigma})}{\partial(\partial_\mu A_\nu)}F_{\alpha\beta}\right].
\end{align*}
Since
\begin{align*}
    \frac{\partial(F_{\alpha\beta})}{\partial(\partial_\mu A_\nu)} = \frac{\partial(\partial_\alpha A_\beta - \partial_\beta A_\alpha)}{\partial(\partial_\mu A_\nu)} = \delta^{\mu}_\alpha\delta^{\nu}_\beta - \delta^{\mu}_\beta\delta^{\nu}_\alpha,
\end{align*}
we get 
\begin{align*}
    &\left(\frac{\partial\mathcal{L}}{\partial(\partial_{\mu}A_\nu)}\right) \\&= -\frac{1}{4}g^{\alpha\lambda}g^{\beta\sigma}[(\delta^{\mu}_\alpha\delta^{\nu}_\beta - \delta^{\mu}_\beta\delta^{\nu}_\alpha)F_{\lambda\sigma} + (\delta^{\mu}_\lambda\delta^{\nu}_\sigma - \delta^{\mu}_\sigma\delta^{\nu}_\lambda)F_{\alpha\beta}]\\
    &= -\frac{1}{4}[F^{\mu\nu}-F^{\nu\mu}+F^{\mu\nu}-F^{\nu\mu}]=-F^{\mu\nu},
\end{align*}
where we have used the fact that $F^{\mu\nu} = -F^{\nu\mu}$ is anti-symmetric. Therefore,
\begin{equation}
    \partial_{\mu}\left(\frac{\partial\mathcal{L}}{\partial(\partial_{\mu}A_\nu)}\right) = -\partial_\mu F^{\mu\nu}.
\end{equation}
And then the dynamic equations of electromagnetic fields are given by
\begin{equation}
    \partial_\mu F^{\mu\nu} = J^{\nu}. \label{DE}
\end{equation}

Going back to SI units, we can explicitly write
\begin{equation}
    F^{\mu\nu} = 
\begin{bmatrix}
0 & -E_x/c & -E_y/c & -E_z/c\\
E_x/c &  0  & -B_z & B_y\\
E_y/c & B_z &  0  & -B_x\\
E_z/c & -B_y & B_x & 0\\
\end{bmatrix},
\end{equation}
and the dynamic equations become
\begin{equation}\label{eq_ft}
    \partial_\mu F^{\mu\nu} = \mu_0 J^{\nu}.
\end{equation}
One can check that equations (\ref{eq_ft}) indeed give half of Maxwell's equations (\ref{columb}) and (\ref{ampere}).That is, the rest equations (\ref{thompson}) and (\ref{faraday}) do not describe the dynamics of electromagnetic fields. They are usually written in a compact form
\begin{subequations}
\begin{align}
&\epsilon_{\mu\nu\rho\sigma}\partial^\nu F^{\rho\sigma}=
\epsilon_{\mu\nu\rho\sigma}\partial^\nu(\partial^\rho A^\sigma-\partial^\sigma A^\rho)=0\\
\Longleftrightarrow &2\epsilon_{\mu\nu\rho\sigma}\partial^\nu\partial^\rho A^\sigma
=-2\epsilon_{\mu\nu\rho\sigma}\partial^\nu\partial^\rho A^\sigma=0, \label{bianchi}
\end{align}
\end{subequations}
where $\epsilon_{\mu\nu\rho\sigma}$ is the Levi-Civita tensor. Eq. (\ref{bianchi}) is known as the Bianchi identity. Bianchi identity are constraints on components of electromagnetic fields rather than their dynamics. 

We now can clearly see that for $\nu = 0,1,2,3$, we get four dynamic equations from (\ref{eq_ft}) (identical to equations (\ref{columb}) and (\ref{ampere})) and four constraints from (\ref{bianchi}) for $\mu = 0,1,2,3$ (identical to equations (\ref{thompson}) and (\ref{faraday})). These equations are actually not independent because only four of the six curl equations from (\ref{faraday}) and (\ref{ampere}) are independent, which can be seen from the two vector identities below:
\begin{align*}
    &\vec{\nabla}\cdot(\vec{\nabla}\times\vec{E}) = 0,\\
    &\vec{\nabla}\cdot(\vec{\nabla}\times\vec{B}) = 0.
\end{align*}
Therefore, there are three independent dynamic equations and three constraints. Electromagnetic fields contain two vector fields in 3D space, while each vector field has three degrees of freedom. Therefore, six degrees of freedom constrained by three independent constraints left three independent degrees of freedom to be determined by the three independent dynamic equations. 

If we take the 4-potential $A_\mu$ as our variables instead of $\vec{E}$ and $\vec{B}$, the constraints from Bianchi identities are already encoded. With a scalar field $\phi$ and a vector field $\vec{A}$ in 3D space, we have $1+3 = 4$ degrees of freedom that can be solved by three independent dynamic equations and one gauge condition. 

The four dynamic equations (\ref{DE}) we mentioned above are reduced from a more general and higher-level theory: Yang-Mills theory. Here we would like to emphasize that the Lorentz covariance of electromagnetism can be seen as required by Yang-Mills equation only in order to keep the invariance of speed of light, while Bianchi identity is generically covariant under an arbitrary linear coordinate transformation, as shown below. 

\subsection{Helmholtz's theorem}

Helmholtz's theorem (see Appendix B of \cite{pollack_electromagnetism_2002}) is the fundamental theorem of vector calculus which states sufficiently smooth, rapidly decaying vector field $\vec{F}$ in three dimensions can be resolved into the sum of an irrotational (curl-free) vector field (also named as longitudinal component in physics) $\vec{F}_l$ and solenoidal (divergence-free) vector field (transverse component) $\vec{F}_t$:
\begin{subequations}\label{decomposition}
\begin{align}
    &\vec{F} = \vec{F}_l + \vec{F}_t,\\
    &\vec{\nabla}\cdot\vec{F}_t = 0,\\
    &\vec{\nabla}\times\vec{F}_l = 0,
\end{align}
\end{subequations}
In momentum space $\vec{F}(\vec{r})\rightarrow\Tilde{\vec{F}}(\vec{q})$, the decomposition has the form:
\begin{subequations}
\begin{align}
    &\Tilde{\vec{F}} = \Tilde{\vec{F}}_l + \Tilde{\vec{F}}_t,\\
    &\vec{q}\cdot\Tilde{\vec{F}}_t = 0,\\
    &\vec{q}\times\Tilde{\vec{F}}_l = 0,
\end{align}
\end{subequations}
which indicates that the transverse component has two non-zero components and the longitudinal component has only one non-zero component. Then it is clear that two of the three degrees of freedom of $\vec{F}$ are associated with its transverse component while the rest one is associated with its longitudinal component. 

As a result of (\ref{decomposition}), the irrotational part can always be written as a gradient of a scalar field (scalar potential) $U$ and the solenoidal part can always be written as a curl of a vector field (vector potential) $\vec{W}$:
\begin{subequations}
\begin{align}
    &\vec{F}_t = \vec{\nabla}\times\vec{W},\\
    &\vec{F}_l = -\vec{\nabla}U,\\
    &\vec{F} = -\vec{\nabla}U + \vec{\nabla}\times\vec{W},
\end{align}
\end{subequations}
provided that
\begin{align*}
    &\vec{\nabla}\cdot\vec{F}_t  = \vec{\nabla}\cdot(\vec{\nabla}\times\vec{W})\equiv 0,\\
    &\vec{\nabla}\cdot\vec{F}_l  = \vec{\nabla}\cdot(-\vec{\nabla}\cdot U)\equiv 0.
\end{align*}
The potentials $U$ and $\vec{W}$ have direct relations respectively with the divergence and curl of $\vec{F}$:
\begin{subequations}
\begin{align}
    &U(\vec{r}) = \int d^3\vec{r}'\frac{\vec{\nabla}\cdot\vec{F}(\vec{r}')}{4\pi\left|\vec{r}-\vec{r}'\right|},\\
    &\vec{W}(\vec{r}) = \int d^3\vec{r}'\frac{\vec{\nabla}\times\vec{F}(\vec{r}')}{4\pi\left|\vec{r}-\vec{r}'\right|}.
\end{align}
\end{subequations}
As a reminder, we have
\begin{align*}
    &\vec{\nabla}\cdot\vec{F} = \vec{\nabla}\cdot\vec{F}_l,\\
    &\vec{\nabla}\times\vec{F} = \vec{\nabla}\times\vec{F}_t.
\end{align*}

In summary, a rapidly decaying vector field can be uniquely determined with both a specified divergence and a specified curl:
\begin{subequations}
\begin{align}
    &\vec{\nabla}\cdot\vec{F} = d,\\
    &\vec{\nabla}\times\vec{F} = \vec{c}.\label{curl_eq}
\end{align}
\end{subequations}
We emphasize once again that only two of the three equations in (\ref{curl_eq}) are independent since one has to impose the divergenceless condition
\begin{equation*}
    \vec{\nabla}\cdot\vec{c} = 0
\end{equation*}
for consistency, because the divergence of a curl is always zero.

\subsection{ Symmetry of the dynamic equations}\label{symmetry}

It is well known that the Maxwell equations are Lorentz covariant, i.e., their form is invariant under the Lorentz transformation.  However, this is in fact the symmetry of the dynamic equations (\ref{DE}). The Bianchi identity is of a much larger symmetry as we will see below.\\
 
 \noindent{\it Dynamic equations}\\

Make a linear transformation on the left hand side of Eq. (\ref{DE}),  
\begin{eqnarray}
\partial'_\mu F'^{\mu\nu}=L^\alpha_\mu\partial_\alpha L^\mu_\beta L^\nu_\gamma F^{\beta\gamma}.
\end{eqnarray}
Generally, this does not go back to the form of Eq. (\ref{DE}).  We then require $L^\alpha_\mu L^\mu_\beta=\delta^\alpha_\beta$,   so that
\begin{eqnarray}
\partial'_\mu F'^{\mu\nu}=L^\alpha_\mu\partial_\alpha L^\mu_\beta L^\nu_\gamma F^{\beta\gamma}=L^\nu_\gamma \partial_\alpha F^{\alpha\gamma}.
\end{eqnarray}
On the other hand, the right hand side of (\ref{DE}) is transformed by $j'^\mu=L^\mu_\nu j^\nu$. Thus, the dynamic equations read
\begin{eqnarray}
\partial'_\mu F'^{\mu\nu}=L^\nu_\gamma \partial_\alpha F^{\alpha\gamma}=L^\nu_\gamma j^\gamma=j'^\nu.
\end{eqnarray}
Therefore, the dynamic equations are covariant under the linear transition $L$. If we require the speed of light is invariant, $L$ is the Lorentz transformation. \\

\noindent{\it Bianchi identity}\\

We denote
\begin{eqnarray}
W_\mu=\epsilon_{\mu\nu\rho\sigma}\partial^\nu F^{\rho\sigma}.
\end{eqnarray}
We do a linear transformation to $W_\mu$ according to the tensor algebra
\begin{eqnarray}
W'_\mu=\epsilon_{\mu\nu\rho\sigma}\partial'^\nu F'^{\rho\sigma},
\end{eqnarray}
where $\partial'^\nu=\Lambda^\nu_\alpha\partial^\alpha$ with  a linear transformation $\Lambda$ which is not restricted to the Lorentz transformation, e.g., the Galileo transformation or an $SO(3)$ rortation and  so on. (Here, the Galileo transformation does not only act on the coordinate but also act on the field. )  The rank-2 tensor is transformed as $F'^{\rho\sigma}=\Lambda^\rho_\alpha\Lambda^\sigma_\beta F^{\alpha\beta}$.  For the Levi-Civita tensor, one has
\begin{subequations}
\begin{align}
&\Lambda_\mu^\alpha\Lambda_\nu^\beta\Lambda_\rho^\gamma\Lambda_\sigma^\delta\epsilon_{\alpha\beta\gamma\delta}=\epsilon_{\mu\nu\rho\sigma},\\
&\Lambda_\nu^\beta\Lambda_\rho^\gamma\Lambda_\sigma^\delta\epsilon_{\xi\beta\gamma\delta}=\delta_\xi^\alpha\Lambda_\nu^\beta\Lambda_\rho^\gamma\Lambda_\sigma^\delta\epsilon_{\alpha\beta\gamma\delta}=(\Lambda^{-1})_\xi^\mu\Lambda_\mu^\alpha\Lambda_\nu^\beta\Lambda_\rho^\gamma\Lambda_\sigma^\delta\epsilon_{\alpha\beta\gamma\delta}=(\Lambda^{-1})_\xi^\mu\epsilon_{\mu\nu\rho\sigma}
\end{align}
\end{subequations}
Hence, 
\begin{align}
\epsilon_{\mu\nu\rho\sigma}\partial'^\nu F'^{\rho\sigma}&=\epsilon_{\mu\nu\rho\sigma}\Lambda^\nu_\alpha\Lambda^\rho_\gamma\Lambda^\sigma_\delta \partial^\alpha F^{\gamma\delta}=(\Lambda^{-1})_\mu^\alpha
\epsilon_{\alpha\nu\rho\sigma}\partial^\nu F^{\rho\sigma}=(\Lambda^{-1})_\mu^\alpha W_\alpha.
\end{align}
That is, $W'_\mu=(\Lambda^{-1})_\mu^\alpha W_\alpha$ as its definition.

This means  that after an arbitrary linear transformation,  the Bianchi identities are form invariant. 

Back to the electric and magnetic fields, the transformed Bianchi identity reads
\begin{equation}
W'_\mu=(\Lambda^{-1})_\mu^0\nabla\cdot \vec B+{\vec \Lambda^{-1}}_\mu\cdot (\nabla\times {\vec E}+\frac{\partial \vec B}{\partial t})=0.
\end{equation}
Namely,
\begin{align}
&\nabla'\cdot \vec B'=(\Lambda^{-1})_0^0\nabla\cdot \vec B+{\vec \Lambda^{-1}}_0\cdot (\nabla\times {\vec E}+\frac{\partial \vec B}{\partial t})=0,\\
&(\nabla'\times \vec E')_i+\frac{\partial B'_i}{\partial t}=(\Lambda^{-1})_i^0\nabla\cdot \vec B+{\vec \Lambda^{-1}}_i\cdot (\nabla\times {\vec E}+\frac{\partial \vec B}{\partial t})=0.
\end{align}
 The transformed Bianchi identity is the linear combination of $\nabla\cdot \vec B$ and $\nabla\times {\vec E}+\frac{\partial \vec B}{\partial t}$.
 When the linear transformation is dependent on the media, all the media parameters are included in $\Lambda$. 
 
 For the Galileo transformation, 
 \begin{subequations}
 \begin{align}
 \Lambda&=\left(\begin{array}{cccc} 1&v_1&v_2&v_3\\
                                                        0&1&0&0\\
                                                        0&0&1&0\\
                                                        0&0&0&1
                                                        \end{array}\right),\\
    \Lambda^{-1}&=\left(\begin{array}{cccc} 1&-v_1&-v_2&-v_3\\
                                                        0&1&0&0\\
                                                        0&0&1&0\\
                                                        0&0&0&1
                                                        \end{array}\right),                       \end{align}                          
                                                        \end{subequations}
 it is easy to see
\begin{eqnarray}
W_\mu'=W_\mu.
\end{eqnarray}
That is
$$\nabla'\cdot \vec B'=\nabla\cdot \vec B,~~\nabla'\times\vec E'+\frac{\partial \vec B}{\partial t'}=\nabla\times\vec E+\frac{\partial \vec B}{\partial t}.$$
Under the Galileo transformation, the Bianchi identity is not only covariant but also invariant.

The conclusions on the symmetry of the Maxwell theory are that

(1) The symmetry of the Bianchi identity is much larger than the dynamic equations.  The symmetry of the electromagnetic theory is Lorentz symmetry. 

(2) There is not any expanded version of the Bianchi identity when we do not make the mistake.
Under the Galileo transformation, the Bianchi identity is not only covariant but also invariant.  

\section{Form invariance of Maxwell's equations under low speed limits}\label{append_trans}
In this appendix, we show explicitly why Maxwell's equations are form invariant after frame transformations under the low speed limits. 

\subsection{Equations of transformations for a special configuration}
Since we always have $v/c$ to appear as a whole, we can safely go to $c=1$ unit for simplicity in notation ($\beta = v/c = v$).

For a special configuration that $\vec{v}$ along $x$ axis and two frames aligned with each other, we get the following transformations:
\begin{subequations}\label{low0}
\begin{align}
    E_x' &= E_x,\\
    E_y' &= E_y - vB_z,\\
    E_z' &= E_z + vB_y,\\
    B_x' &= B_x,\\
    B_y' &= B_y + vE_z,\\
    B_z' &= B_z - vE_y,
\end{align}
\end{subequations}
\begin{subequations}\label{low3}
\begin{align}
    x' &= x-vt,\\
    y' &= y,\\
    z' &= z,\\
    t' &= t-vx,
\end{align}
\end{subequations}
\begin{subequations}\label{low4}
\begin{align}
    J_x' &= J_x-v\rho,\\
    J_y' &= J_y,\\
    J_z' &= J_z,\\
    \rho' &= \rho-vJ_x,
\end{align}
\end{subequations}
\begin{subequations}\label{difftrans}
\begin{align}
    &\frac{\partial}{\partial x'} = \frac{\partial}{\partial x} + v\frac{\partial}{\partial t},\\
    &\frac{\partial}{\partial y'} = \frac{\partial}{\partial y},\\
    &\frac{\partial}{\partial z'} = \frac{\partial}{\partial z},\\
    &\frac{\partial}{\partial t'} = \frac{\partial}{\partial t} + v\frac{\partial}{\partial x},
\end{align}
\end{subequations}
and
\begin{subequations}\label{invdifftrans}
\begin{align}
    &\frac{\partial}{\partial x} = \frac{\partial}{\partial x'} - v\frac{\partial}{\partial t'},\\
    &\frac{\partial}{\partial y} = \frac{\partial}{\partial y'},\\
    &\frac{\partial}{\partial z} = \frac{\partial}{\partial z'},\\
    &\frac{\partial}{\partial t} = \frac{\partial}{\partial t'} - v\frac{\partial}{\partial x'}.
\end{align}
\end{subequations}

Although our frames are now in a special configuration, we do not lose any generality in our discussion since physics laws are independent of the orientations of coordinate systems.

\subsection{Frame transformation of Maxwell's equations under low speed limits}
With every piece of elements prepared, we now transform the four Maxwell's equations in the primed frame:
\begin{subequations}\label{pmaxwell}
\begin{align}
    \vec{\nabla}'\cdot\Vec{E'} &= \rho'/\epsilon_0,\label{m1}\\
    \vec{\nabla}'\cdot\Vec{B'} &= 0,\label{m2}\\
    \vec{\nabla}'\times\Vec{E'} &= -\frac{\partial}{\partial t'}\Vec{B'},\label{m3}\\
    \vec{\nabla}'\times\Vec{B'} &= \mu_0\Vec{J'} + \frac{1}{c^2}\frac{\partial}{\partial t'}\Vec{E'},\label{m4}
\end{align}
\end{subequations}
to the unprimed frame, one at a time. In the above equations we retain the speed of light $c$ to avoid possible confusions, but we will go back to $\mu_0\epsilon_0 = 1/c^2 = 1$ unit in below.

\subsubsection{Transformation of $\vec{\nabla}'\cdot\Vec{E'} = \rho'/\epsilon_0$}\label{t1}
We firstly write it out in Cartesian coordinates:
\begin{equation}
    \frac{\partial E_x'}{\partial x'} + \frac{\partial E_y'}{\partial y'} + \frac{\partial E_z'}{\partial z'} = \rho'/\epsilon_0,
\end{equation}
then express the primed quantities with unprimed quantites based on transformations (\ref{difftrans}a), (\ref{low0}) and (\ref{low4}d):

\begin{equation}\label{sub1}
    (\frac{\partial }{\partial x} + v\frac{\partial }{\partial t}) E_x+ \frac{\partial}{\partial y}(E_y-vB_z) + \frac{\partial}{\partial z}(E_z+vB_y) = (\rho-vJ_x)/\epsilon_0,
\end{equation}

Although this looks unlike the original (\ref{m1}), we argue that the terms that are at the same order on both sides should be equal separately, since $v$ is arbitrary. Therefore, we get:
\begin{align*}
    &\frac{\partial E_x}{\partial x} + \frac{\partial E_y}{\partial y} + \frac{\partial E_z}{\partial z} = \rho/\epsilon_0,\\
    &\frac{\partial B_z}{\partial y} - \frac{\partial B_y}{\partial z} = J_x/\epsilon_0 + \frac{\partial E_x}{\partial t}.
\end{align*}
The first line is equation (\ref{m1}) transformed to the unprimed frame, while the second line is nothing special but the $x$ component of equation (\ref{m4}) transformed to the unprimed frame (which will be seen again in below), if we recall $\epsilon_0\mu_0 =1/c^2 = 1$ so that $1/\epsilon_0 = 1/c^2\epsilon_0 = \mu_0$.

Just in case some readers do not like the above argument, we can instead use equations (\ref{low0}e), (\ref{low0}f) and (\ref{low4}a) to replace $B_y$, $B_z$ and $J_x$ with $B_y'$, $B_z'$, $E_y$, $E_z$,$J_x'$ and $\rho$ in equation (\ref{sub1}), and replace some differential operators:

\begin{align*}
    &(\frac{\partial E_x}{\partial x} + v\frac{\partial E_x}{\partial t}) + \frac{\partial}{\partial y}(E_y-v(B_z'+vE_y)) + \frac{\partial}{\partial z}(E_z+v(B_y'-vE_z)) = (\rho-v(J_x'+v\rho))/\epsilon_0,\\
    \Longrightarrow{}&(\frac{\partial E_x}{\partial x} + v(\frac{\partial }{\partial t'}-v\frac{\partial}{\partial x'})E_x) + \frac{\partial}{\partial y}(E_y-v(B_z'+vE_y)) + \frac{\partial}{\partial z}(E_z+v(B_y'-vE_z)) = (\rho-v(J_x'+v\rho))/\epsilon_0
\end{align*}

After omitting higher order terms in $v^2$, we realize that some terms in the above equation cancel according to (\ref{m4}):
\begin{small}
\begin{equation*}
    v(\frac{\partial E_x}{\partial t'} - \frac{\partial B_z'}{\partial y} + \frac{\partial B_y'}{\partial z}) = v(\frac{\partial E_x'}{\partial t'} - \frac{\partial B_z'}{\partial y'} + \frac{\partial B_y'}{\partial z'}) = -vJ_x'/\epsilon_0,
\end{equation*}
\end{small}
then the rest is precisely
\begin{equation}
    \frac{\partial E_x}{\partial x} + \frac{\partial E_y}{\partial y} + \frac{\partial E_z}{\partial z} = \rho/\epsilon_0.
\end{equation}
We prove that equation (\ref{m1}) is form invariant to the frame transformation in low speed limits.

\subsubsection{Transformation of $\vec{\nabla}'\cdot\Vec{B'} = 0$} \label{t2}
Again, write it out in Cartesian coordinates:
\begin{equation}
    \frac{\partial B_x'}{\partial x'} + \frac{\partial B_y'}{\partial y'} + \frac{\partial B_z'}{\partial z'} = 0,
\end{equation}
then substitute relevant quantities:
\begin{equation}
    (\frac{\partial }{\partial x} + v\frac{\partial }{\partial t}) B_x+ \frac{\partial}{\partial y}(B_y+vE_z) + \frac{\partial}{\partial z}(B_z-vE_y) = 0,
\end{equation}
which again looks unlike (\ref{m2}), but can be separated into two equations by their orders according to the same argument above:
\begin{align*}
    &\frac{\partial B_x}{\partial x} + \frac{\partial B_y}{\partial y} + \frac{\partial B_z}{\partial z} = 0,\\
    &\frac{\partial E_z}{\partial y} - \frac{\partial E_y}{\partial z} = -\frac{\partial B_x}{\partial t},
\end{align*}
where the second line above is the $x$ component of equation (\ref{m3}) transformed to the unprimed frame (will be seen again below). The readers can also use the same trick we did for the first Maxwell's equation to convince themselves that the same result
\begin{equation}
    \frac{\partial B_x}{\partial x} + \frac{\partial B_y}{\partial y} + \frac{\partial B_z}{\partial z} = 0
\end{equation}
can be got after omitting higher order terms.

\subsubsection{Transformation of $\vec{\nabla}'\times\Vec{E'} = -\partial\Vec{B'}/\partial t'$}\label{t3}
This is a equation with three components, we write them out in Cartesian coordinates:

\begin{equation}
    \left(\frac{\partial E_z'}{\partial y'} - \frac{\partial E_y'}{\partial z'},\frac{\partial E_x'}{\partial z'} - \frac{\partial E_z'}{\partial x'},\frac{\partial E_y'}{\partial x'} - \frac{\partial E_x'}{\partial y'}\right)  = \left(-\frac{\partial B_x'}{\partial t'},-\frac{\partial B_y'}{\partial t'},-\frac{\partial B_z'}{\partial t'}\right)
\end{equation}

There are actually three equations instead of one. We substitute the relevant quantities and write the three equations separately:

\begin{align*}
    \frac{\partial }{\partial y}(E_z+vB_y) - \frac{\partial }{\partial z}(E_y-vB_z) = -(\frac{\partial }{\partial t} + v\frac{\partial}{\partial x})B_x,\\
    \frac{\partial }{\partial z}E_x - (\frac{\partial }{\partial x} + v\frac{\partial}{\partial t})(E_z+vB_y) = -(\frac{\partial }{\partial t} + v\frac{\partial}{\partial x})(B_y+vE_z),\\
    (\frac{\partial }{\partial x} + v\frac{\partial}{\partial t})(E_y-vB_z)  - \frac{\partial}{\partial y} E_x= -(\frac{\partial }{\partial t} + v\frac{\partial}{\partial x})(B_z-vE_y).
\end{align*}

All the second order terms appear in the above equations should be omitted, and each of the above equations can be separated into two equations by their orders:
\begin{align*}
    \frac{\partial E_z}{\partial y} - \frac{\partial E_y}{\partial z} = -\frac{\partial B_x}{\partial t},\\
    \frac{\partial B_x}{\partial x} + \frac{\partial B_y}{\partial y} + \frac{\partial B_z}{\partial z} = 0,\\
    \frac{\partial E_x}{\partial z} - \frac{\partial E_z}{\partial x} = -\frac{\partial B_y}{\partial t},\\
    -\frac{\partial E_z}{\partial t} - \frac{\partial B_y}{\partial x} = -\frac{\partial E_z}{\partial t} - \frac{\partial B_y}{\partial x},\\
    \frac{\partial E_y}{\partial x}  - \frac{\partial E_x}{\partial y} = -\frac{\partial B_z}{\partial t},\\
    -\frac{\partial B_z}{\partial x} + \frac{\partial E_y}{\partial t} = -\frac{\partial B_z}{\partial x} + \frac{\partial E_y}{\partial t}.
\end{align*}
We can clearly see that the first, third and fifth lines are together the equation (\ref{m3}) transformed to the unprimed frame. The second line is the transformed (\ref{m2}) that was got above, while the rest two lines are identities.

The readers can perform the same trick as for the first Maxwell's equation to get the same result:

\begin{equation}
    \left(\frac{\partial E_z}{\partial y} - \frac{\partial E_y}{\partial z},\frac{\partial E_x}{\partial z} - \frac{\partial E_z}{\partial x},\frac{\partial E_y}{\partial x} - \frac{\partial E_x}{\partial y}\right)  = \left(-\frac{\partial B_x}{\partial t},-\frac{\partial B_y}{\partial t},-\frac{\partial B_z}{\partial t}\right)
\end{equation}

\subsubsection{Transformation of $\vec{\nabla}'\times\Vec{B'} = \mu_0\Vec{J'} + \frac{1}{c^2}\partial \Vec{E'}/\partial t'$}\label{t4}
The three components written in Cartesian coordinates are:

\begin{equation}
    \left(\frac{\partial B_z'}{\partial y'} - \frac{\partial B_y'}{\partial z'},\frac{\partial B_x'}{\partial z'} - \frac{\partial B_z'}{\partial x'},\frac{\partial B_y'}{\partial x'} - \frac{\partial B_x'}{\partial y'}\right) = \left(\mu_0J_x' + \frac{\partial E_x'}{\partial t'},\mu_0J_y' + \frac{\partial E_y'}{\partial t'},\mu_0J_z' + \frac{\partial E_z'}{\partial t'}\right)
\end{equation}

Just like in \ref{t3}, we substitute the relevant quantities and write the three equations separately:

\begin{align*}
    \frac{\partial }{\partial y}(B_z-vE_y) - \frac{\partial }{\partial z}(B_y+vE_z) &= \mu_0(J_x-v\rho) + (\frac{\partial }{\partial t} + v\frac{\partial}{\partial x})E_x,\\
    \frac{\partial }{\partial z}B_x - (\frac{\partial }{\partial x} + v\frac{\partial}{\partial t})(B_z-vE_y) &= \mu_0J_y + (\frac{\partial }{\partial t} + v\frac{\partial}{\partial x})(E_y-vB_z),\\
    (\frac{\partial }{\partial x} + v\frac{\partial}{\partial t})(B_y + vE_z)  - \frac{\partial}{\partial y} B_x &= \mu_0J_z + (\frac{\partial }{\partial t} + v\frac{\partial}{\partial x})(E_z+vB_y).
\end{align*}

Following the same strategy in \ref{t3}, we get six separate equations:
\begin{align*}
    \frac{\partial B_z}{\partial y} - \frac{\partial B_y}{\partial z} = \mu_0J_x + \frac{\partial E_x}{\partial t},\\
    -\frac{\partial E_y}{\partial y} - \frac{\partial E_z}{\partial z} = -\mu_0\rho + \frac{\partial E_x}{\partial x},\\
    \frac{\partial B_x}{\partial z} - \frac{\partial B_z}{\partial x} = \mu_0J_y + \frac{\partial E_y}{\partial t},\\
    \frac{\partial E_y}{\partial x} - \frac{\partial B_z}{\partial t} = -\frac{\partial B_z}{\partial t} + \frac{\partial E_y}{\partial x},\\
    \frac{\partial B_y}{\partial x}  - \frac{\partial B_x}{\partial y} = \mu_0J_z + \frac{\partial E_z}{\partial t},\\
    \frac{\partial B_y}{\partial t} + \frac{\partial E_z}{\partial x} = \frac{\partial B_y}{\partial t} + \frac{\partial E_z}{\partial x}.
\end{align*}
Like in \ref{t3}, the first, third and fifth line are together the equation (\ref{m4}) transformed to the unprimed frame. The second line is the transformed (\ref{m1}) that was got above (if we recall that $\mu_0 = \mu_0c^2 = 1/\epsilon_0$), while the rest two lines are identities.
Once again, the readers can perform the same trick as for the first Maxwell's equation to get the same result:

\begin{equation}
    \left(\frac{\partial B_z}{\partial y} - \frac{\partial B_y}{\partial z},\frac{\partial B_x}{\partial z} - \frac{\partial B_z}{\partial x},\frac{\partial B_y}{\partial x} - \frac{\partial B_x}{\partial y}\right) = \left(\mu_0J_x + \frac{\partial E_x}{\partial t},\mu_0J_y + \frac{\partial E_y}{\partial t},\mu_0J_z + \frac{\partial E_z}{\partial t}\right).
\end{equation}

In summary, in the derivation of \ref{t1}, \ref{t2}, \ref{t3} and \ref{t4}, we get nothing else but the following four equations and some identities:
\begin{subequations}\label{tmaxwell}
\begin{align}
    \vec{\nabla}\cdot\Vec{E} &= \rho/\epsilon_0,\label{tm1}\\
    \vec{\nabla}\cdot\Vec{B} &= 0,\label{tm2}\\
    \vec{\nabla}\times\Vec{E} &= -\frac{\partial}{\partial t}\Vec{B},\label{tm3}\\
    \vec{\nabla}\times\Vec{B} &= \mu_0\Vec{J} + \frac{1}{c^2}\frac{\partial}{\partial t}\Vec{E}.\label{tm4}
\end{align}
\end{subequations}
These four equations express relationships between electromagnetic fields and the charge/currents, which are sufficient to determine the distribution of fields in the unprimed frame together with proper boundary and initial conditions (we've put $c$ back for here and below). In conclusion, equations (\ref{tmaxwell}) are Maxwell's equations in the unprimed frame, which has the same form as those in the primed frame. Therefore, Maxwell's equations are form invariant under the low speed limits of Lorentz's transformation.

\section{The speed of light and Doppler effect}\label{doppler}

It is well known that the formation of Lorentz's transformation and the theory of special relativity is to retain the fact that the speed of light is always $c$ regardless of the relative motion between frames, or more particular, the light source and the observer. Here we want to ensure its first order approximation in low speed limits still preserves this feature.

There are two ways to investigate the problem. The first one is more based on logic: we directly look at the Maxwell's equations in both frames: since (\ref{maxwell}) and (\ref{tmaxwell}) have the same form, especially the same constants $\epsilon_0$, $\mu_0$ and $1/c^2 = \epsilon_0\mu_0$, the wave equations derived from them must also have the same form with same constants in it. Therefore, if a plane wave solution of (\ref{maxwell}) propagates at speed $c$ in the primed frame, a plane wave solution of (\ref{tmaxwell}) also propagates at speed of $c$ in the unprimed frame.

The second perspective is to look at a wave solution. Let's take a simple case: just look at a solution in the primed frame in a region in absence of media:
\begin{subequations}
\begin{align}
    \vec{E'}(\vec{r'},t') &= E_0'f(x'-ct')\hat{y'},\\
    \vec{B'}(\vec{r'},t') &= B_0'f(x'-ct')\hat{z'},
\end{align}
\end{subequations}
where $B_0' = E_0'/c$.
Transform it to the unprimed frame using (\ref{low3}) and the inverse of (\ref{low0}):

\begin{align*}
    \vec{E}(\vec{r},t) &= (E_0'+vB_0')f((x-vt)-c(t-vx/c^2))\hat{y} = E_0'(1+\frac{v}{c})f(x(1+\frac{v}{c})-ct(1+\frac{v}{c}))\hat{y},\\
    \vec{B}(\vec{r},t) &= (B_0'+vE_0'/c^2)f((x-vt)-c(t-vx/c^2))\hat{z} = B_0'(1+\frac{v}{c})f(x(1+\frac{v}{c})-ct(1+\frac{v}{c}))\hat{z}.
\end{align*}

It is seen that the last equality in both of the above equations can be simplified by defining another function $g(\cdot) = f((1+v/c)\cdot)$, so that the wave observed in the unprimed frame is:
\begin{subequations}
\begin{align}
    \vec{E}(\vec{r},t) &= E_0'(1+\frac{v}{c})g(x-ct)\hat{y},\\
    \vec{B}(\vec{r},t) &= B_0'(1+\frac{v}{c})g(x-ct)\hat{z},
\end{align}
\end{subequations}
which is still a wave travelling in speed $c$.

Additionally, the Doppler effect can also be seen. Suppose the wave solution we are looking at in the primed frame is monochromatic, that is
\begin{equation}
    f(x'-ct') = \cos(k'x'-\omega' t'),
\end{equation}
where $\omega'/k' = c$then the wave observed in the unprimed system becomes
\begin{equation}
    f[(1+\frac{v}{c})(x-ct)] = \cos[(k'x-\omega't)(1+\frac{v}{c})].
\end{equation}
It can be read out that the frequency observed in the unprimed frame is $1+v/c$ times larger than that in the primed frame, which is the first order approximation of the exact Doppler factor $\sqrt{(1+v/c)/(1-v/c)}$. For more information about Doppler's effect, see Steven Weinberg's \textit{Gravitation and Cosmology} Chapter 2.2.

\bibliographystyle{apsrev4-1} 
\bibliography{reference} 

\end{document}